\documentclass[
    ,final            
  ]
  {aipproc}

\layoutstyle{6x9}
\newcommand{\be}{\begin{eqnarray}}
\newcommand{\ben}{\begin{eqnarray}\nonumber}
\newcommand{\ee}{\end{eqnarray}}
\newcommand{\nee}{\nonumber \end{eqnarray}}

\begin{document}

\title{Phase Transition to Exact Susy}

\classification{<95.36.+x,98.80.-k,12.90.+b,12.60.Jv>}
\keywords      {<string landscape, anthropic principle,
susy phase transition,dark energy>}

\author{L. Clavelli}{
  address={Dept. of Physics and Astronomy\\
University of Alabama\\
Tuscaloosa AL 35487
}}

\begin{abstract}
  The anthropic principle is based on the observation that, within
narrow bounds, the laws of physics are such as to have allowed the
evolution of life.  The string theoretic approach to understanding
this observation is based on the expectation that the effective
potential has an enormous number of local minima with  
different particle masses and perhaps totally different fundamental
couplings and space time topology.  The vast
majority of these alternative universes are totally inhospitable to
life, having, for example, vacuum energies near the natural (Planck) 
scale.  The statistics, however, are assumed to be such that a few of 
these local minima (and not more) have a low enough vacuum energy and 
suitable other properties to support life.  In the inflationary era,
the ``multiverse" made successive transitions between the available
minima until arriving at our current state of low vacuum energy.
String theory, however, also suggests that the absolute minimum of
the effective potential is exactly supersymmetric.  Questions then
arise as to why the inflationary era did not end by a transition to
one of these, when will the universe make the phase transition to the
exactly supersymmetric ground state, and what will be the properties
of this final state. 
\end{abstract}

\maketitle

   There is no doubt that the fact that we are here puts constraints on
the laws of physics.  The question is whether this provides a sort
of explanation for the way things are.  The string landscape scenario
attempts to provide a statistical understanding for the
anthropic principle.  This understanding depends on the assumption
that the effective potential contains at most a few alternative 
universes in which life could evolve.   There is some debate as to
whether or not such viable universes are truly rare \cite{weakless}.

   The current universe with broken supersymmetry seems to be
accelerating outwards due to a positive vacuum energy density
\be
         \epsilon = 3560\, MeV/m^3 = (.0023\, eV)^4 \quad .
\label{vacenergy}
\ee
The natural value that might have been expected for this quantity is
\be
      M_{Planck}^4 = 10^{127} \,MeV/m^3 \quad 
\ee
some $124$ orders of magnitude greater than observed.

\begin{figure}
  \includegraphics[height=.3\textheight]{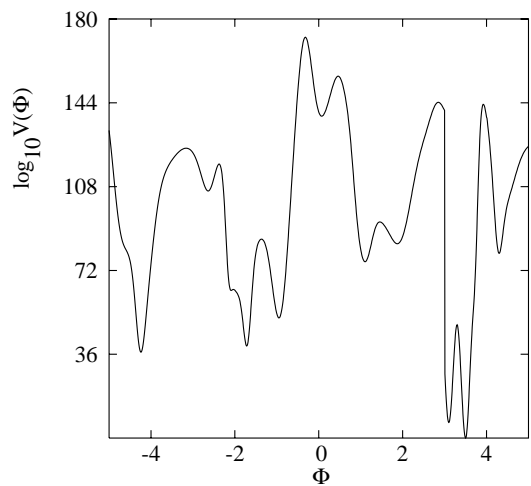}
\caption{A schematic representation of the effective potential in the string landscape
picture. Our world with a small vacuum energy is shown together with the neighboring
exact susy phase with zero vacuum energy.  The y axis scale is broken and taken to be linear at small $V$.}
\label{landscape}
\end{figure}

String theory suggests that, in addition to our broken susy universe,  
there is a lower lying neighboring
valley in the string landscape described by a perfect supersymmetry (susy)
\cite{Giddings}
\cite{future} and, most likely, a vanishing cosmological constant as pictured in
figure \ref{landscape}.  
Some of the prominent features of the exact susy phase are independent of
the exact space-time topology as long as the cosmological constant is
not much greater in absolute value than our current one.  
We expect that this susy minimum is the true vacuum and, therefore, the 
final phase of the universe.  While our primary interest, at present, is in
the final transition from our broken susy world to the exact susy universe,
it is thought that the inflationary phase in the very early universe 
corresponded to a sequence of
similar phase transitions to progressively lower vacuum energies.  Many such
scenarios have been considered recently by Susskind and collaborators 
\cite{Susskind} as well as by others.  It is crucial for the rise of life 
that the universe escaped from the inflationary phase to a phase of low
vacuum energy \cite{Weinberg} but also that this transition occured slowly but 
soon enough that the universe had not been ripped apart beforehand by inflation.

     We have proposed \cite{future} that the primary distinguishing property
of matter in the exact susy phase relative to our universe 
is an effective weakening of the Pauli Principle.   This is due to the fact
that,
in the broken susy world, every atom above helium is characterized by energy
permanently stored in a Pauli tower of electrons and in a separate tower of
nucleons in the atomic nucleus.  In exact susy, conversion of fermion pairs to
degenerate scalar pairs not governed by the Pauli principle allows the release
of this energy:
\be
     f f \rightarrow {\tilde f}{\tilde f} \quad .
\label{pairconversion}
\ee
This process \cite{CP} occurs in every susy model with or without $R$ parity violation. Thus, following a phase transition
to exact susy, fermions in excited states will convert in pairs to bosons 
which can then drop into the ground state.

A phase transition in vacuum will begin with the nucleation of a bubble of 
true vacuum with radius greater than a critical radius depending on the 
surface tension $S$.
\be
       R_c = \frac{3S}{\epsilon}  \quad .
\ee
Although a supersymmetric true vacuum was not
specifically considered,  it was generically predicted
\cite{Coleman} that such a bubble will expand in the vacuum at the 
speed of light converting all matter in its path to the new phase.  
Although there can be no advance warning of the arrival time 
of a susy bubble nucleated in the vacuum,
the inevitability of such a phase change is implied if
the effective potential of string theory is dynamically determined and
the true vacuum is supersymmetric.  

The four basic questions posed in ref.\,\cite{future} are
\begin{enumerate}
\item{\bf Could life have arisen if there had been a phase transition 
directly from the inflationary era to the exact susy minimum?}
 
 There are several tentative arguments that no such possibility exists \cite{future}. For example, one could note that galactic evolution seems to rely on a
large dark matter component to provide the gravitational well within which
normal matter can condense into galaxies.  In an exact susy world the lightest
susy particle would not serve this function.  Other sources of dark matter are,
of course, possible.

\item{\bf Could life survive, or re-establish itself, following a transition
from our broken susy world to the exact susy world?}
 
If it is confirmed that the rise of life
would have been impossible if there had been a direct transition from an inflationary era to an exact susy universe, one could still
ask whether an exact susy universe could support life if there was an
intermediate
broken susy phase.  Like the time critical property of the transition from the
inflationary era to our calm broken susy universe, the transition to exact susy 
might also be time critical.  If the current accelerating phase lasts too long,  
most stars will consist of white dwarfs out of causal contact with each other.
At that point
it is unlikely that life could be revived through a susy phase transition.
On the other hand, if the transition takes place
while there are still earth-like planets orbiting burning stars, it is 
conceivable that life could re-establish itself as discussed in point 3 below.  

\item{\bf What would be the primary characteristics of the physics (and biology, if any) of the exactly supersymmetric phase?}
 
The primary distinguishing features of bulk susy matter relative to matter in the
broken susy phase are the greater numbers of states due to supersymmetry and the
weakening of the Pauli Principle due to the possibility of pair conversion from
fermions to bosons according to eq.\,\ref{pairconversion}.  Whenever, in the
broken susy phase, bound fermions are forced into elevated energy levels, in the
susy phase it will be advantageous for them to convert in pairs into their
degenerate susy partners which, being bosons, can drop into the ground state.
Susy atoms will, therefore, consist of zero, one, or two fermionic electrons
but possibly many selectrons.  The entire ground state leptonic cloud will be in
the 1S state.  This has the effect of making susy atoms much smaller in general
than their broken susy counterparts. Smaller atoms in a solution will be 
expected to have slower reaction rates due to the decreased probability of 
collisions but might bind more tightly into molecules because of the smaller
intra-molecular distances.
     
      Assuming degenerate
susy multiplets have the same masses as the standard model particles in
the broken susy world, the atomic weight of snuclei increases rapidly with
atomic number so that stable elements above susy oxygen must have atomic weights
well above $238$.  Since in the broken susy world there are long-lived elements
with atomic weights only up to this number, after a susy phase transition
only elements up to susy oxygen would be expected to be abundant.  
The elements with higher
atomic number would beta decay down to oxygen and below due to Coulomb repulsion
and the absence of an effective Pauli principle \cite{future}.
It is plausible that molecular binding
qualitatively similar to that of our world would then occur.
Since all the elements needed to form DNA and $96\%$ by weight of animal 
species are no heavier than oxygen, 
evolution might be expected to recur leading to the re-emergence of species 
qualitatively similar to many of those in the broken susy world and defined by
the same genetic codes.

\item{\bf Can we estimate the probable time remaining before our universe converts to a susy world?}
 
The vacuum decay probability per unit time per unit volume depends on the vacuum energy of the current phase, eq.\,\ref{vacenergy}. 
Thus the transition rate is proportional to the volume in which a phase change is
possible, proportional in turn to the cube of the scale factor in the
Friedman-Robertson-Walker (FRW) metric which, for positive cosmological constant, is
exponentially growing at large times. 

The natural time scale for the growth in volume of the universe in a FRW metric with
vacuum energy density $\epsilon$ is
\be
    \gamma^{-1} = \frac{1}{\sqrt{24\pi G_N \epsilon}}= 5.61 \cdot 10^9 {\displaystyle {yr}} \quad .
\ee
Depending on the parameters, there is a non-negligible probability 
that the Earth will be swallowed 
by a susy bubble in a time $T$ from today that is smaller than $1/\gamma$.
\be
     P(T) \approx e^{\gamma T} - 1 \quad .
\ee

\end{enumerate}

We have outlined a possible new end-phase scenario for the universe.
A more detailed review of this scenario is also available
\cite{valley}.

\begin{theacknowledgments}
  This work was partially supported by the DOE under grant number
DE-FG02-96ER-40967.  We acknowledge stimulating discussions with
Paul Cox, Irina Perevalova, Tim Lovorn, and Stephen Barr.
\end{theacknowledgments}

\end{document}